\documentclass[letterpaper,english,aps]{revtex4}
\usepackage[T1]{fontenc}
\usepackage[latin1]{inputenc}
\usepackage{amsmath}
\usepackage{graphicx}
\usepackage{amssymb}

\makeatletter
 \newenvironment{lyxlist}[1]
   {\begin{list}{}
     {\settowidth{\labelwidth}{#1}
      \setlength{\leftmargin}{\labelwidth}
      \addtolength{\leftmargin}{\labelsep}
      }}
   {\end{list}}

\usepackage{amssymb}

\usepackage{amssymb}

\usepackage{graphicx}
\usepackage{geometry}

\usepackage{babel}

\usepackage{babel}

\usepackage{babel}
\makeatother
\begin{document}

\title{Thermoelectric and Thermomagnetic Effects in Dilute Plasmas}

\author{L. S. García-Colín}

\address{Departamento de Física, Universidad Autónoma Metropolitana-Iztapalapa,
Av. Purísima y Michoacán S/N, México D. F. 09340, México. Also at
El Colegio Nacional, Luis González Obregón 23, Centro Histórico, México
D. F. 06020, México.}

\author{A. L. García-Perciante}

\address{Depto. de Matemáticas Aplicadas y Sistemas, Universidad Autónoma
Metropolitana-Cuajimalpa, Av. Pedro A. de los Santos No. 84, México
D. F, México}

\author{A. Sandoval-Villalbazo}

\address{Departamento de Física y Matemáticas, Universidad Iberoamericana,
Prolongación Paseo de la Reforma 880, México D. F. 01210, México.}

\date{\today{}}

\begin{abstract}
When an electrically charged system is subjected to the action of
an electromagnetic field, it responds by generating an electrical
current. In the case of a multicomponent plasma other effects, the
so called cross effects, influence the flow of charge as well as the
heat flow. In this paper we discuss these effects and their corresponding
transport coefficients in a fully ionized plasma using Boltzmann's
equation. Applications to non-confined plasmas, specially to those
prevailing in astrophysical systems are highlighted. Also, a detailed
comparison is given with other available results.
\end{abstract}
\maketitle

\section{Introduction}

In this paper we address ourselves to the study of the behavior of
a fully ionized dilute plasma which is subjected to the action of
an electromagnetic field. Although the primary response of the plasma
to an electric field is the generation of an electrical current, if
a magnetic field is also present other effects arise including a heat
current. This gives rise, according to the tenets of classical irreversible
thermodynamics (CIT) to a number of interesting and important cross
effects which, to our knowledge, have received little or no attention
in the literature when examined under the framework of microscopic
equations. This is the main objective of our work, to use the well
known Boltzmann equation to calculate the explicit forms of the flux-force
relations demanded by CIT and provide explicit expressions for the
ensuing transport coefficients as functions of the density, temperature
and magnetic field. We believe that the results will turn out to be
valuable for many astrophysical systems and also other situations
related to non-confined plasmas.

To achieve this program, in Section II we summarize the kinetic basics
of the problem using standard techniques of the kinetic theory of
gases, in Section III we establish all flux-force relations appearing
in the system including the explicit calculation of the transport
coefficients, and in Section IV we present a critical discussion of
our results including a comparison with others available in the literature.

\section{Kinetic Background}

Foundations of the methods here applied are available in the somewhat
extensive literature on the subject
\cite{key-one,key-two,key-three,key-four,key-five,key-six,key-seven,key-eight}
so we shall avoid detailed algebraic steps which the reader may easily
find in any of the works here cited.

We consider a dilute binary mixture of charged particles with masses
$m_{a}$, $m_{b}$; charges $e_{a}=-e_{b}=e$ taking $Z=1$ merely
for didactic purposes. The number density $n=n_{a}+n_{b}$ and the
mass density $\rho=m_{a}n_{a}+m_{b}n_{b}$ where $n_{i}$ is the number
density of species $i$ ($i=a,\, b$). The fields $\vec{E}$ and $\vec{B}$
acting on the system are the self-consistent fields as determined
from Maxwell's equations and the possibility of having an additional
external magnetic field is taken care off by defining the total field
$\vec{B}_{T}=\vec{B}+\vec{B}^{(\text{ext})}$. These fields are included
in Boltzmann's equation through the Lorentz force \cite{key-nine}
so that if $f_{i}\left(\vec{r},\,\vec{v}_{i},\, t\right)$ is the
single particle distribution function for species $i$ this equation
reads as \begin{equation}
\frac{\partial f_{i}}{\partial t}+\vec{v_{i}}\cdot\frac{\partial f_{i}}{\partial\vec{r}}+
\frac{1}{m_{i}}\left(\vec{F}_{i}+e_{i}\vec{v_{i}}\times\vec{B}\right)\cdot
\frac{\partial f_{i}}{\partial\vec{v_{i}}}=\sum_{j=a}^{b}J\left(f_{i}f_{j}\right)\,.\label{1}\end{equation}
Here, the electric field is included as part of the total conservative
force.\begin{equation}
\vec{F}_{i}=\vec{F}_{i}^{(\text{ext})}+e_{i}\vec{E}\label{2}\end{equation}
and\begin{equation}
J\left(f_{i}f_{j}\right)\equiv\int...\int\left\{ f_{i}
\left(\vec{v}_{i}'\right)f_{j}\left(\vec{v}_{j}'\right)-f_{i}\left(\vec{v}_{i}
\right)f_{j}\left(\vec{v}_{j}\right)\right\} \sigma\left(\vec{v}_{i},\,\vec{v}_{j}
\rightarrow\vec{v}_{i}',\,\vec{v}_{j}'\right)d\vec{v}_{j}\, d\vec{v}_{i}'\, d\vec{v}_{j}'\label{3}
\end{equation}
where the primes denote the velocities after a collision $\left(\vec{v}_{i},\,\vec{v}_{j}\rightarrow
\vec{v}_{i}',\,\vec{v}_{j}'\right)$
and the probability that this collision occurs satisfies microscopic
reversibility namely,\begin{equation}
\sigma\left(\vec{v}_{i},\,\vec{v}_{j}\rightarrow\vec{v}_{i}',\,\vec{v}_{j}'\right)=\sigma
\left(\vec{v}_{i}',\,\vec{v}_{j}'\rightarrow\vec{v}_{i},\,\vec{v}_{j}\right)\qquad i,\, j=a,\, b\label{4}
\end{equation}
In words, Eq.\,(\ref{4}) guarantees the existence of inverse collisions
among the two species.

For our purposes, we concentrate on a general feature of Eq.\,(\ref{1}),
namely the derivation of the conservation equations for the chosen
local variables: $\rho_{i}\left(\vec{r},\, t\right)$ the local mass
density for species $i$, $\rho\vec{u}\left(\vec{r},\, t\right)$
the local baricentric momentum per unit volume and the local internal
energy density $\varepsilon\left(\vec{r},\, t\right)$. In fact multiplication
of Eq.\,(\ref{1}) by $m_{i}$, $m_{i}\vec{v}_{i}$ and $\frac{1}{2}m_{i}v_{i}^{2}$
and integration over $\vec{v}_{i}$ leads effortlessly to the following
results:

i) Mass conservation equation,\begin{equation}
\frac{\partial\rho_{i}}{\partial t}+
\nabla\cdot\left(\rho_{i}\vec{u}\right)=-\nabla\cdot\vec{J}_{i}\label{5}\end{equation}
where $\vec{u}$, the barycentric velocity, is defined
as\begin{equation} \rho\,\vec{u}\left(\vec{r},\,
t\right)=\sum_{i=a}^{b}\rho_{i}\vec{u}_{i}\left(\vec{r},\,
t\right)\label{6}\end{equation} with\begin{equation}
\vec{u}_{i}\left(\vec{r},\, t\right)= \frac{1}{n_{i}}\int
f_{i}\left(\vec{r},\,\vec{v}_{i},\,
t\right)\vec{v}_{i}d\vec{v}_{i} \equiv\left\langle
\vec{v}_{i}\right\rangle \label{7}\end{equation} and
clearly\begin{equation} \left\langle \psi\right\rangle
=\frac{1}{n_{i}}\int f_{i}\left(\vec{r},\,\vec{v}_{i},\,
t\right)\psi\left(\vec{v}_{i}\right)d\vec{v}_{i}\label{14}\end{equation}
The diffusive flux $\vec{J}_{i}$ is\begin{equation}
\vec{J}_{i}=m_{i}\int f_{i}\left(\vec{r},\,\vec{v}_{i},\,
t\right)\vec{c}_{i}d\vec{v}_{i}=m_{i}n_{i}\left\langle
\vec{c}_{i}\right\rangle \label{8}\end{equation} and the thermal
or random velocity $\vec{c}_{i}$ is simply given by
$\vec{c}_{i}=\vec{v}_{i}-\vec{u}\left(\vec{r},\, t\right)$. From
these definitions it follows readily that
$\vec{J}_{a}+\vec{J}_{b}=\rho_{a}\left\langle
\vec{c}_{a}\right\rangle +\rho_{b}\left\langle
\vec{c}_{b}\right\rangle =0$.

ii) Momentum conservation equation,\begin{equation}
\frac{\partial}{\partial t}\left(\rho\vec{u}\right)+\nabla\cdot\left(\tau^{k}+
\rho\vec{u}\vec{u}\right)=\sum_{i}n_{i}\vec{F}_{i}+\vec{J}_{T}\times\vec{B}\label{9}\end{equation}
where the total charge current $\vec{J}_{T}$ is defined as
\begin{equation}
\vec{J}_{T}=\sum_{i=a}^{b}n_{i}e_{i}\left\langle \vec{v}_{i}\right\rangle \label{10}
\end{equation}
and the stress tensor is given by\begin{equation}
\tau^{k}=\sum_{i=a}^{b}m_{i}\int f_{i}\left(\vec{r},\,\vec{v}_{i},\, t\right)\vec{c}_{i}\vec{c}_{i}d\vec{v}_{i}
\label{11}\end{equation}
Notice that according to Eq.\,(\ref{11}) $\tau^{k}$ is a symmetric
tensor. Also, in Eq.\,(\ref{9}) $\vec{F}_{i}$ is the total conservative
force acting on species $i$ including the electric field $\vec{E}$.

iiI): Internal energy equation,
\begin{equation}
\rho\frac{\partial}{\partial t}\left(\frac{\varepsilon}{\rho}\right)+\nabla\cdot\vec{J}_{Q}+\tau^{k}:
\nabla\vec{u}-\vec{J}_{c}\cdot\vec{E}'=0\label{12}
\end{equation}
where\begin{equation}
\rho\,\varepsilon\left(\vec{r},\, t\right)=\sum_{i}\frac{1}{2}\rho_{i}\left\langle c_{i}^{2}\right\rangle \label{13}
\end{equation}
The heat flux $\vec{J}_{Q}$ is given by
\begin{equation}
\vec{J}_{Q}=\sum_{i}\frac{1}{2}\rho_{i}\left\langle \vec{c}_{i}c_{i}^{2}\right\rangle \label{15}
\end{equation}
and the conductive current $\vec{J}_{c}$ is defined as\begin{equation}
\vec{J}_{c}=\sum_{i}n_{i}e_{i}\left\langle \vec{c}_{i}\right\rangle \label{16}\end{equation}
Finally, $\vec{E}'=\vec{E}+\vec{u}\times\vec{B}$ is the {}``effective\char`\"{}
electric field that is, the field measured by an observer moving with
the baricentric velocity $\vec{u}$.

It is important to underline the fact that the set of conservation
equations (\ref{5}), (\ref{9}) and (\ref{12}) is incomplete. Indeed
we have six independent variables and six equations, but all the currents
are unknown, $\vec{J}_{i}$, $\vec{J}_{c}$, $\vec{J}_{Q}$ and $\tau^{k}$.
In order to derive a complete set of equations we need to express
these currents as functions of the local state variables which implies
clearly that we must solve Eq.\,(\ref{1}) for $f_{i}\left(\vec{r},\,\vec{v}_{i},\, t\right)$
and somehow introduce the local variables in the solution.

But there is another ingredient that we must require form our solutions.
We want to write the two important currents namely $\vec{J}_{Q}$
and the conduction current $\vec{J}_{c}$ in terms of the gradients
in the system. This implies seeking relations such as
\begin{eqnarray}
\vec{J}_{Q} & = & -\kappa\nabla T-\mathcal{T}\,\nabla\phi\nonumber \\
\vec{J}_{c} & = & -\mathcal{B}\nabla T-\sigma\nabla\phi\label{17}
\end{eqnarray}
since $\vec{E}=-\nabla\phi$. These are the linear flux-force relations
as demanded by CIT where the transport coefficients $\kappa$, $\sigma$,
$\mathcal{T}$ and $\mathcal{B}$ have to be determined. Here $\kappa$
and $\sigma$ are the ordinary thermal and electrical conductivities
whereas $\mathcal{T}$ and $\mathcal{B}$ are the so called Thomson's
and Benedicks coefficients respectively. Their form will be radically
modified by the presence of a magnetic field contrary to the case
where $\vec{B}=0$ and only an electrical field is present. Notice
also that from the definition of $\vec{J}_{a}$, the fact that
$\vec{J}_{a}+\vec{J}_{b}=0$
and the definition of $\vec{J}_{c}$,
\begin{equation}
\vec{J}_{c}=\frac{m_{a}+m_{b}}{m_{a}m_{b}}e_{a}\vec{J}_{a}\label{18}
\end{equation}
so the mass and conduction currents are proportional among each other.
We shall not pursue diffusive effects here as they have been reported
elsewhere \cite{key-ten}.

\section{The Transport Coefficients and Linear Laws}

To achieve the program outlined in the previous section, we proceed
with the solution of Boltzmann's equation. The first stage of this
task is standard in the literature
\cite{key-one,key-two,key-three,key-four,key-five,key-six,key-seven,key-eight}
so we merely outline the main steps. One assumes that the single particle
distribution function may be expanded in a power series of the Knudsen
parameter $\epsilon$ which is a measure of the spatial gradients
of the local variables in the system. This series is taken using the
local Maxwellian distribution as a reference state since this function
is, as well known, a solution to the homogeneous term, the collisional
term in Eq.\,(\ref{1}). Moreover this expansion holds for times
much longer than the collision times so that the distribution function
is assumed to be a time independent functional of the conserved densities.
Known in the literature as the Hilbert-Chapman-Enskog expansion it
thus reads as\begin{equation}
f_{i}\left(\vec{r},\,\vec{v_{i}},\, t\right)=f_{i}^{\left(0\right)}
\left(\vec{r},\,\vec{v_{i}}|\right)\left[1+\epsilon\varphi_{i}^{\left(1\right)}
\left(\vec{r},\,\vec{v_{i}}|\right)+...\right]\,,\label{19}\end{equation}
here $f_{i}^{\left(0\right)}$ is the local Maxwellian distribution
function and the dash in all functions in the right side of Eq.\,(\ref{19})
implies the time dependence through $n_{i}\left(\vec{r},\, t\right)$,
$\vec{u}\left(\vec{r},\, t\right)$ and $\varepsilon\left(\vec{r},\, t\right)$.

After Eq.\,(\ref{19}) is substituted back into Eq.\,(\ref{1})
and recalling that the zeroth order in $\epsilon$ term is just
$\sum_{j=a}^{b}J\left(f_{i}f_{j}\right)=0$,
whose solution is $f_{i}^{(0)}$, one obtains the linearized Boltzmann
equations for $\varphi_{i}^{\left(1\right)}$ whose solutions are
known to be of the form
\begin{equation}
\varphi_{i}^{\left(1\right)}=\mathbb{B}_{i}:\nabla\vec{u}+\vec{\mathbb{A}}_{i}\cdot\nabla\ln T+
\vec{\mathbb{D}}_{i}\cdot\vec{d}_{ij}\qquad i,\, j=a,\, b\label{20}
\end{equation}
Equations (\ref{20}), the first order in $\epsilon$ solutions to
Eq.\,(\ref{1}) characterize the well known Navier-Stokes-Fourier
regime of magnetohydrodynamics. In Eq.\,(\ref{20})
\begin{equation}
\vec{d}_{ij}\equiv\nabla\left(\frac{n_{i}}{n}\right)+
\frac{n_{i}n_{j}}{n}\left(\frac{m_{j}-m_{i}}{\rho}\right)
\frac{\nabla p}{p}+\frac{n_{i}n_{j}}{\rho p}\left(m_{i}e_{j}-m_{j}e_{i}\right)
\vec{E}'=-\vec{d}_{ji}\,,\label{21}
\end{equation}
where $p=nkT$ is the local pressure and the temperature is introduced
through the standard ideal gas relationship, namely,
$\varepsilon\left(\vec{r},\, t\right)=\frac{3}{2}kT\left(\vec{r},\, t\right)$.
The vector $\vec{d}_{ij}$ is called the diffusive force and contains
two main contributions, the first two terms related to ordinary and
pressure diffusion and the third and fourth terms, the diffusive effect
arising from the of electromagnetic fields. Also, in this expression
we have assumed that no external forces are acting on the plasmas,
$\vec{F}_{i}^{(\text{ext})}=0,\, i=a,\, b$.

In what follows we shall ignore the tensorial term $\mathbb{B}_{i}:\nabla\vec{u}$
since by Curie's principle \cite{key-eleven}, it will not couple
with first rank tensors such as $\nabla T$ and $\vec{E}'$. Also,
the structure of the vectors $\vec{\mathbb{A}}_{i}$ and $\vec{\mathbb{D}}_{i}$
and is not arbitrary. They must be linear combinations of the independent
vectors available, $\vec{c}_{i}$, $\vec{c}_{i}\times\vec{B}$ and
$\left(\vec{c}_{i}\times\vec{B}\right)\times\vec{B}$. Therefore,
\begin{equation}
\vec{\mathbb{A}}_{i}=\mathbb{A}_{i}^{\left(1\right)}\vec{c}_{i}+
\mathbb{A}_{i}^{\left(2\right)}\vec{c}_{i}\times\vec{B}+
\mathbb{A}_{i}^{\left(3\right)}\left(\vec{c}_{i}\times\vec{B}\right)\times\vec{B}\,,\label{22}
\end{equation}
and a similar expression for $\vec{\mathbb{D}}_{i}$. The scalar functions
$\mathbb{A}_{i}^{\left(1\right)}$, $\mathbb{A}_{i}^{\left(2\right)}$
and $\mathbb{A}_{i}^{\left(3\right)}$may depend on all scalar quantities
such as $c_{i}^{2}$, $B^{2}$, $\left(\vec{c}_{i}\cdot\vec{B}\right)^{2}$,
$n_{i}$, $T$ and so on. The next step is to expand these functions
in terms of a complete set of functions namely, the Sonine polynomials.
For instance,
\begin{equation}
\mathbb{A}_{i}^{(1)}=\sum_{m=0}^{\infty}a_{i}^{(1)(m)}S_{3/2}^{(m)}\left(c_{i}^{2}\right)\,,\label{23}
\end{equation}
and similar expressions for $\mathbb{A}_{i}^{\left(2\right)}$, etc.
Use of Eqs.\,(\ref{22}) and (\ref{23}) in Eq.\,(\ref{20}) allows
a formal computation of the fluxes in the system. Let us assume that
$m_{b}\gg m_{a}$, $m_{b}$ and $m_{a}$ associated with the mass
of a proton and an electron, respectively, in a hydrogen plasma. Since
then, $\vec{J}_{c}=\frac{e}{m_{a}}\vec{J}_{a}$ we find, using Eqs.\,(\ref{8}),
(\ref{20}), (\ref{22}), and (\ref{23}) as well as the orthogonality
conditions of Sonine polynomials that
\begin{eqnarray}
\vec{J}_{c} & = & \frac{en_{a}k}{m_{a}}\left\{ a_{a}^{(1)(0)}\nabla T+a_{a}^{(2)(0)}\nabla T\times\vec{B}+
a_{a}^{(3)(0)}\left(\vec{B}\cdot\nabla T\right)\vec{B}\right.\nonumber \\
 &  & \left.+T\left[d_{a}^{(1)(0)}\vec{d}_{ab}^{(e)}+d_{a}^{(2)(0)}\vec{d}_{ab}^{(e)}\times\vec{B}+
 \left(\vec{B}\cdot\vec{d}_{ab}^{(e)}\right)\vec{B}\, d_{a}^{(3)(0)}\right]\right\} \,,\label{24}
 \end{eqnarray}
This result is quite interesting. Out of the infinite number of
coefficients required to characterize the scalar functions
$\mathbb{A}_{a}^{(j)}$, $\mathbb{D}_{a}^{(j)}$ ($j=1,\,2,\,3$) as
illustrated by Eq.\,(\ref{23}), only one coefficient is necessary
for each of these functions in order to compute the conduction
current. These coefficients can be obtained from the solution of
the integral equations that such functions must obey and whose
solutions will not be discussed here. The procedure to obtain them
is, once more, a standard technique in kinetic theory and we refer
the reader to the available literature if he wishes to purpose the
details
\cite{key-one,key-eight,key-two,key-three,key-four,key-five,key-six,key-seven}.
In Eqs.\,(\ref{24}) $\vec{d}_{ab}^{(e)}$ is the electromagnetic
contribution of the diffusive force
\begin{equation}
\vec{d}_{ab}^{(e)}=\frac{n_{a}n_{b}\left(m_{a}+m_{b}\right)e}{p\rho}\vec{E}'\label{25}
\end{equation}

To obtain the expression for the heat flux we must first recall a
result from CIT. For a multicomponent system the standard definition
of heat flux is \cite{key-eleven}
\begin{equation}
\vec{J}'_{Q}=\vec{J}_{Q}-\sum_{r}h_{r}\frac{\vec{J}_{r}}{m_{r}}\label{26}
\end{equation}
where $h_{r}$ is the enthalpy of the $r^{th}$ component which is
equal to $\frac{5}{2}kT$ for an ideal gas. In kinetic theory the
analog of this equation is clearly given by
\begin{equation}
\vec{J}'_{Q}=kT\sum_{i=a}^{b}\int\left(\frac{m_{i}c_{i}^{2}}{2kT}-
\frac{5}{2}\right)\vec{c}_{i}f_{i}^{(0)}\varphi_{i}^{\left(1\right)}d\vec{c}_{i}\label{27}
\end{equation}
where use has been made of Eqs.\,(\ref{8}), (\ref{15}), and (\ref{19})
recalling that $\vec{J}'_{Q}=0$ in the local equilibrium state.

Now, when Eq.\,(\ref{20}) is substituted in Eq.\,(\ref{27}) the
term $\vec{\mathbb{A}}_{i}\cdot\nabla T$ gives rise to the ordinary
Fourier heat conduction. With all its modalities arising from the
presence of a magnetic field, this has been thoroughly studied in
a separate paper \cite{key-twelve} so we shall not deal with it here.
However the term from Eq.\,(\ref{25}) of our interest here, is the
one giving rise to the electromagnetic influence on the heat current
and reads as,
\begin{equation}
\vec{J}_{Q}^{(e)}=\frac{5}{2}\left(kT\right)^{2}\sum_{i=a}^{b}\frac{n_{i}}{m_{i}}
\left(\vec{d}_{ab}^{(e)}d_{i}^{(1)(1)}
+d_{i}^{(2)(1)}\vec{B}\times\vec{d}_{ab}^{(e)}+d_{i}^{(3)(1)}
\left(\vec{B}\cdot\vec{d}_{ab}^{(e)}\right)\vec{B}\right)\label{28}
\end{equation}
To analyze Eqs.\,(\ref{24}) and (\ref{28}) we consider a cartesian
coordinate system and let $\vec{B}$ point along the $z$-axis, $\vec{B}=B\hat{k}$,
$\hat{k}$ being the unit vector. Then for any scalar $N$, we define,
\begin{equation}
\nabla_{\parallel}N\equiv\frac{\partial N}{\partial z}\quad;
\quad\nabla_{\perp}N\equiv\frac{\partial N}{\partial x}\hat{i}+\frac{\partial N}
{\partial y}\hat{j}\quad;\quad\nabla_{s}N\equiv\frac{\partial N}{\partial x}\hat{j}-
\frac{\partial N}{\partial y}\hat{i}\label{29}
\end{equation}
the last definition corresponding to a vector perpendicular to both
the parallel and perpendicular components, respectively. Let us now
apply Eq.\,(\ref{29}) to our currents. The thermal contribution
to the conduction current reads
\begin{equation}
\vec{J}_{c}^{(T)}=\frac{n_{a}ke}{m_{a}}\left\{ \left(a_{a}^{(1)(0)}+
B^{2}a_{a}^{(3)(0)}\right)\nabla_{\parallel}T+a_{a}^{(1)(0)}\nabla_{\perp}T+
B\, a_{a}^{(2)(0)}\nabla_{s}T\right\} \label{30}
\end{equation}

This is the first important result in this paper, since it
contains all the thermoelectrical effects in the plasma when a
magnetic field is present. Notice that if $\vec{B}=0$
\begin{equation} \vec{J}_{c}^{(T)}=-\mathcal{T}\,\nabla
T\label{31}
\end{equation}
as required by CIT. $\mathcal{T}$ is the
standard Thompson thermoelectric coefficient which turns out to be
given by
\begin{equation}
\mathcal{T}=5.334\frac{n_{a}ke}{m_{e}}\tau\label{32}
\end{equation}
where, $\tau$, the mean collision time calculated from the
collision integrals \cite{key-eight,key-ten} is shown to be given
by
\begin{equation}
\tau=\frac{24\pi^{3/2}\epsilon_{0}^{2}\sqrt{m_{e}}}{n\psi
e^{4}}\left(kT\right)^{3/2}\label{33}
\end{equation} and $\epsilon_{0}=8.554\times10^{-12}F/m$ is the electrical constant.
In the presence of a magnetic field there are, besides the current
parallel to the $z$-axis unaffected by the field, two additional
contributions. Firstly a current along $\nabla_{\perp}T$,
perpendicular to $\vec{B}$, with a \char`\"{}perpendicular Thomson
coefficient\char`\"{}
\begin{equation}
\mathcal{T}_{\perp}=\frac{n_{a}ke\tau}{m_{a}\Delta_{1}\left(x\right)}
\left(-3.1\times10^{-4}x^{4}+68x^{2}+8.53\right)\label{34}
\end{equation}
where $x=\omega\tau$, $\omega=\frac{eB}{m_{e}}$, Larmor's frequency,
$\tau$ is given by Eq.\,(\ref{33}), $\psi$ is the so-called logarithmic
function
\begin{equation}
\psi=\ln\left[1+\left(\frac{16\pi kT\epsilon_{0}}{e^{2}}\lambda_{D}\right)^{2}\right]\label{35}
\end{equation}
and $\lambda_{D}$ is Debye's length defined by
\begin{equation}
\lambda_{D}=\left(\frac{\epsilon_{0}kT}{ne^{2}}\right)^{1/2}\label{36}
\end{equation}
In Eq.\,(\ref{34}) we also used the result that
\begin{equation}
a_{a}^{(1)(0)}=\frac{\tau}{\Delta_{1}\left(x\right)}\left(-3.1\times10^{-4}x^{4}+68x^{2}+8.53\right)\label{37}
\end{equation}
with
\begin{equation}
\Delta_{1}\left(x\right)=4.7\times10^{-4}x^{6}+100x^{4}+32x^{2}+1.6\label{38}
\end{equation}
as obtained from the solution of the integral equations
\cite{key-eight,key-ten}.

Finally, there is a third current in a direction both perpendicular
to $\vec{B}$ and to $\nabla_{\perp}T$ which resembles a \char`\"{}thermal
Hall effect\char`\"{}. The corresponding transport coefficient is
\begin{equation}
\mathcal{T}_{s}=\frac{n_{a}ek}{m_{a}}B\, a_{a}^{(2)(0)}\label{39}
\end{equation}
with
\begin{equation}
B\, a_{a}^{(2)(0)}=\frac{\tau}{\Delta_{1}\left(x\right)}\left(7.8\times10^{-5}x^{3}+17x\right)\label{40}
\end{equation}
To get a clear idea of how these effects compare to each other, the
three coefficients are plotted in Fig. 1 as functions of $x$. Emphasis
must be made on the fact that $x$ is a function of $n$, $B$ and
$T$ so that it must be handled with care when seeking orders of magnitude.

Next we examine the influence of the effective electromagnetic field
on the conduction current. Using the decomposition given in Eq.\,(\ref{29})
one may write that
\begin{equation}
\vec{J}_{c}^{\,\,\left(e\right)}=\frac{e\, n_{a}kT}{m_{a}}\left\{ \left(d_{a}^{(1)(0)}+B^{2}d_{a}^{(3)(0)}\right)
\Gamma E_{z}\hat{k}+d_{a}^{(1)(0)}\left(\vec{d}_{ab}^{\,\,\left(e\right)}\right)_{\perp}+B\, d_{a}^{(2)(0)}
\left(\vec{d}_{ab}^{\,\,\left(e\right)}\right)_{s}\right\} \label{41}
\end{equation}
where\begin{equation}
\begin{array}{c}
\left(\vec{d}_{ab}^{\,\,\left(e\right)}\right)_{\perp}=\left[\left(E_{x}+u_{y}B\right)\hat{i}+
\left(E_{y}-u_{x}B\right)\hat{j}\right]\Gamma\\
\\\left(\vec{d}_{ab}^{\,\,\left(e\right)}\right)_{s}=\left[\left(E_{x}+u_{y}B\right)\hat{j}-\left(E_{y}-u_{x}B\right)
\hat{i}\right]\Gamma\end{array}\label{42}
\end{equation}
and
\begin{equation}
\Gamma=\frac{n_{a}n_{b}\left(m_{a}+m_{b}\right)e}{p\rho}\label{43}
\end{equation}

Using now the fact that $m_{b}\ll m_{a}=m_{e}$, $p=nkT$, the
definition of $\rho$ and assuming a fully ionized plasma so that
$n_{a}=n_{b}=\frac{n}{2}$, Eq.\,(\ref{41}) may be rewritten
as,
\begin{equation}
\vec{J}_{c}^{\,\,\left(e\right)}=\sigma_{\parallel}E_{z}\hat{k}+\sigma_{\perp}
\left(\vec{d}_{ab}^{\,\,\left(e\right)}
\right)_{\perp}+\sigma_{s}\left(\vec{d}_{ab}^{\,\,\left(e\right)}\right)_{s}\label{44}
\end{equation}
which is the second important result in this paper. First notice
that if $B=0$, from Eq.\,(\ref{44}),
$\sigma_{\parallel}=\sigma_{\perp}=\sigma$ and since
$\vec{E}=-\nabla\phi$\begin{equation}
\vec{J}_{c}^{(e)}=-\sigma_{\parallel}\nabla\phi\label{45}\end{equation}
This is the well known form for Ohm's law where\begin{equation}
\sigma_{\parallel}=\frac{e\Gamma
n_{a}kT}{m_{a}}d_{a}^{(1)(0)}\label{46}\end{equation} which, under
the conditions mentioned above and using the value of
$d_{a}^{(1)(0)}=4.31\tau$ extracted from the solution to the
integral equations, gives
\begin{equation}
\sigma_{\parallel}=\frac{25.92\pi^{3/2}\epsilon_{0}^{2}}{\sqrt{m_{a}}e^{2}\psi}
\left(kT\right)^{3/2}\label{47}
\end{equation}
This is the ordinary electrical conductivity as, aside from minor
numerical differences, first derived by Spitzer
\cite{key-thirteen,key-fourteen,key-fifteen} and later by
Braginski \cite{key-sixteen} and Balescu \cite{key-nine}. The
transversal electrical conductivity, when the value for
$d_{a}^{(2)(0)}$ is used, turns out to be
\begin{equation}
\sigma_{\perp}=\frac{6\pi^{3/2}\epsilon_{0}^{2}\left(kT\right)^{3/2}}{\sqrt{m_{a}}e^{2}\psi}
\left(\frac{9.52\times10^{4}x^{4}+20.6x^{2}-13}{\Delta_{2}\left(x\right)}\right)\label{48}
\end{equation}
and $\sigma_{s}$, which corresponds to a \char`\"{}Hall like
effect\char`\"{}, turns out to be
\begin{equation}
\sigma_{s}=\frac{6\pi^{3/2}\epsilon_{0}^{2}\left(kT\right)^{3/2}}{\sqrt{m_{a}}e^{2}\psi}
\left(\frac{0.109x^{5}+2.18\times10^{4}x^{3}+965x}{\Delta_{2}\left(x\right)}\right)\label{50}
\end{equation}
where
\begin{equation}
\Delta_{2}\left(x\right)=1.61x^{6}+3.5\times10^{5}x^{4}+1.65\times10^{4}x^{2}+3.01\label{51}
\end{equation}

To appreciate the order of magnitude of these coefficients they are
plotted in Fig. 2 as functions of $x$. Moreover it is worth pointing
out that from Eq.\,(\ref{48}) it is easily verified that $\lim_{B\rightarrow0}\sigma_{\perp}=\sigma_{\parallel}$
and $\lim_{B\rightarrow0}\sigma_{s}=0$, as seen from Eq.\,(\ref{50}).

The last item we want to deal here with concerns the influence of
$\vec{E}'$ on the heat current which, according to Eqs.\,(\ref{28}),
(\ref{29}) and (\ref{42}), is given by
\begin{equation}
\vec{J}_{Q}^{\,\,\left(e\right)}=\frac{5}{2}\left(kT\right)^{2}\sum_{i=a}^{b}\frac{n_{i}}{m_{i}}\Gamma
\left[\left(d_{i}^{(1)(1)}+B^{2}d_{i}^{(1)(3)}\right)E_{z}\hat{k}+
d_{i}^{(1)(1)}\left(\vec{d}_{ab}^{\,\,\left(e\right)}\right)_{\perp}+
Bd_{i}^{(2)(1)}\left(\vec{d}_{ab}^{\,\,\left(e\right)}\right)_{s}\right]\label{52}
\end{equation}
If $\vec{B}=0$ and assuming a fully ionized hydrogen plasma, using
the value for $d_{i}^{(1)(1)}$ obtained from the solution to the
integral equations one finds that\begin{equation}
\vec{J}_{Q}^{(e)}=-\mathcal{B}_{\parallel}\nabla\phi\label{53}\end{equation}
where\begin{equation}
\mathcal{B}_{\parallel}=\frac{15.9\pi^{3/2}\epsilon_{0}^{2}m_{e}}{e^{2}\psi}\left(\frac{kT}{m_{e}}
\right)^{5/2}\label{54}
\end{equation}
the so-called electropyrosis or Benedicks effect \cite{key-seventeen}.
When $\vec{B}\neq0$ and using the results for the coefficients $d_{i}^{(1)(1)}$
and $d_{i}^{(2)(1)}$ one obtains for the two other coefficients,
namely
\begin{equation}
\mathcal{B}_{\perp}=\frac{15\pi^{3/2}\epsilon_{0}^{2}m_{e}}{e^{2}\psi}
\left(\frac{kT}{m_{e}}\right)^{5/2}\frac{5.04\times10^{-3}x^{4}+1.09\times10^{3}x^{2}+3.2}
{\Delta_{2}\left(x\right)}\label{55}
\end{equation}
and $\lim_{B\rightarrow0}\mathcal{B}_{\perp}=\mathcal{B}_{\parallel}$
and also
\begin{equation}
\mathcal{B}_{s}=\frac{15\pi^{3/2}\epsilon_{0}^{2}m_{e}}{e^{2}\psi}\left(\frac{kT}{m_{e}}\right)^{5/2}
\frac{0.0295x^{3}+222x}{\Delta_{2}\left(x\right)}\label{56}
\end{equation}
$\Delta_{2}\left(x\right)$ defined in Eq.\,(\ref{51}). This effect
has been systematically ignored in the study of transport processes
in plasmas. The importance of the three coefficients is exhibited
in Fig. 3.

\section{discussion of the results }

The results obtained for the nine transport coefficients derived
in the previous section is important for several reasons, mainly
because with one single exception, they have been ignored in the
study of transport processes in plasmas. Let us begin with the
thermoelectric or Thomson coefficients. To our knowledge the only
other derivation of these quantities is that given by Balescu in
his excellent monograph in the subject \cite{key-nine}.
Nevertheless his derivation starts from Landau's version of
Boltzmann's equation which is a Fokker-Planck type equation
\cite{key-eighteen}. This implies in words that transport
processes in plasmas may be visualized as diffusive processes.
Moreover the solution of that kinetic equation is carried out
using a moment-like expansion \textit{a la} Grad which is
different in spirit from the Hilbert-Chapman-Enskog expansion
\cite{key-nineteen,key-twenty}. Here we use the latter to solve
the full Boltzmann equation, which in turn is not a diffusive like
approximation. In Fig. 4 we compare our results for
$\mathcal{T}_{\perp}$ and $\mathcal{T}_{s}$ with those of Balescu
showing explicitly that although qualitatively similar there are
differences in the two methods.

For the electrical conductivity the story is quite different. As
mentioned earlier, it was first derived by Spitzer et al in 1950
\cite{key-thirteen} using a diffusion equation and later by
Spitzer and Härm \cite{key-fourteen} from a Fokker-Planck equation
similar to the one used in his previous paper, but making a mild
attempt to obtain the linear-constitutive laws. In both cases they
obtain, for $\vec{B}=0$, the $\left(kT\right)^{3/2}$ dependence as
in Eq.\,(\ref{47}) with minor numerical differences in the
respective coefficients. Later, Balescu repeated this calculation
methodically \cite{key-nine} and obtained the three conductivities
for the $\vec{B}\neq0$ case.

Finally, the Benedicks coefficients are completely new, to our knowledge
they have never been reported elsewhere. This leads us to the final
remark in this work, namely, when formulating magnetohydrodynamic
equations it is convenient to assess if, in the range of densities,
temperatures and values of the magnetic fields, these cross effects
may be neglected. Otherwise important consequences will be absent
in the ensuing results.

\newpage
\underbar{\large Figure Captions}{\large \par}

\begin{lyxlist}{00.00.0000}
\item [Fig.\,1.\:]The values of the three thermoelectric
coefficients for $n=10^{21}cm^{-3}$ and $T=10^{7}K$. The solid
line is $\mathcal{T}_{\parallel}$, the dashed line
$\mathcal{T}_{\perp}$ and the dotted line $\mathcal{T}_{s}$. \item
[Fig.\,2.\:]The values of the three electrical conductivities for
$n=10^{21}cm^{-3}$ and $T=10^{7}K$. The solid line is
$\sigma_{\parallel}$, the dashed line $\sigma_{\perp}$ and the
dotted line $\sigma_{s}$. Notice the predominance of $\sigma_{s}$
over $\sigma_{\parallel}$ for a certain range of values of $x$.
\item [Fig.\,3.\:]The Benedicks coefficients for
$n=10^{21}cm^{-3}$ and $T=10^{7}K$. The solid line is
$\mathcal{B}_{\parallel}$, the dashed line $\mathcal{B}_{\perp}$
and the dotted line $\mathcal{B}_{s}$. \item [Fig.\,4.\:]The ratio
$\mathcal{T}_{\perp}/T_{\parallel}$ (dashed) and
$\mathcal{T}_{S}/T_{\parallel}$ (dotted) is plotted (right) and
compared with the corresponding Balescu's results (left), as in
Ref. \cite{key-nine}.
\end{lyxlist}
\newpage

\begin{figure}
\includegraphics[width=3in,height=3in]{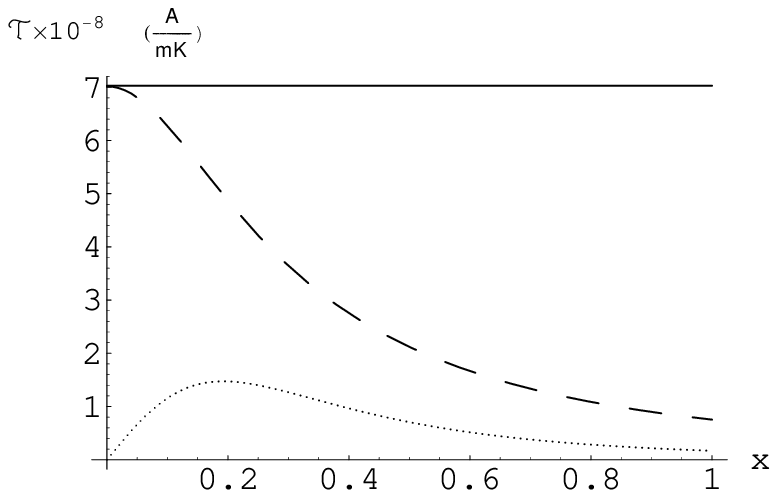}
Fig. 1
\label{fig:F1}
\end{figure}
\newpage

\begin{figure}
\includegraphics[width=3in,height=3in]{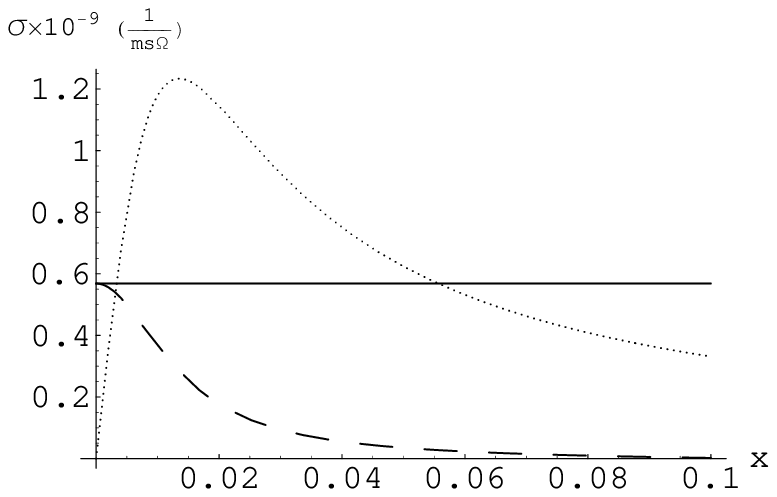}
Fig. 2
\label{fig:F2}
\end{figure}

\newpage

\begin{figure}
\includegraphics[width=3in,height=3in]{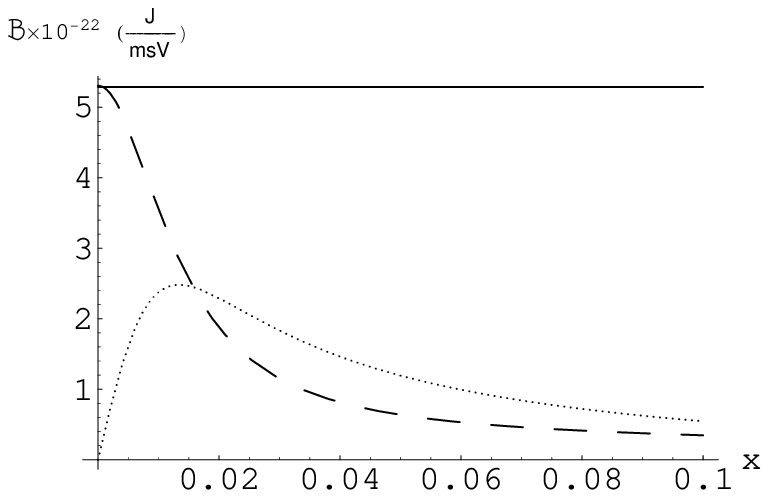}
Fig. 3
\label{fig:F3}
\end{figure}

\newpage
\begin{figure}
\includegraphics[width=3in,height=3in]{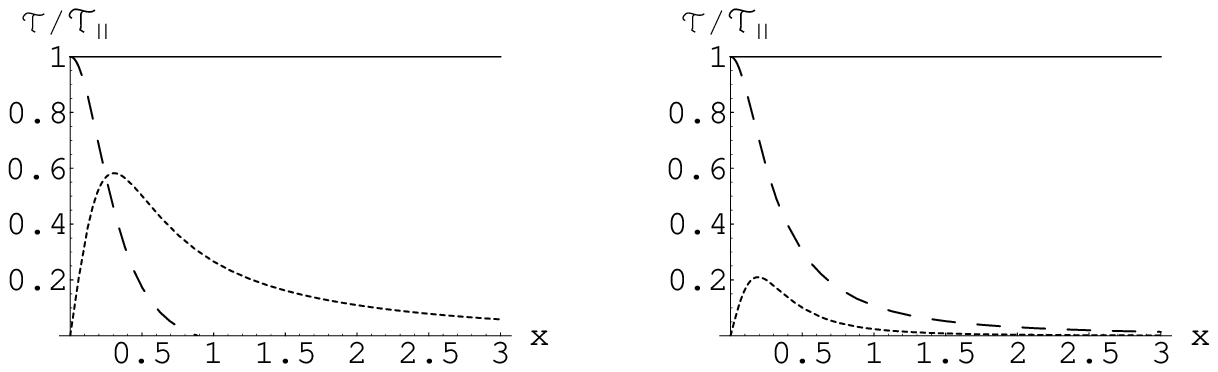}
Fig. 4
\label{fig:F4}
\end{figure}

\end{document}